# Quantifying the Morphologies and Dynamical Evolution of Galaxy Clusters. I. The Method

David A. Buote[1]

Massachusetts Institute of Technology

John C. Tsai[2]

NASA/Ames Research Center

## ABSTRACT

We describe and test a method to quantitatively classify clusters of galaxies according to their projected morphologies. This method will be subsequently used to place constraints on cosmological parameters ($\Omega$ and the power spectrum of primordial fluctuations on scales at or slightly smaller than that of clusters) and to test theories of cluster formation. We specifically address structure that is easily discernible in projection and dynamically important to the cluster. The method is derived from the two-dimensional multipole expansion of the projected gravitational potential and yields dimensionless *power ratios* as morphological statistics. If the projected mass profile is used to characterize the cluster morphology, the power ratios are directly related to the cluster potential. However, since detailed mass profiles currently exist for only a few clusters, we use the X-ray–emitting gas as an alternative tracer of cluster morphology. In this case, the relation of the power ratios to the potential is qualitatively preserved. We demonstrate the feasibility of the method by analyzing simulated observations of simple models of X-ray clusters using the instrument parameters of the ROSAT PSPC. For illustrative purposes, we apply the method to ROSAT PSPC images of A85, A514, A1750, and A2029. These clusters, which differ substantially in their X-ray morphologies, are easily distinguished by their respective power ratios. We discuss the suitability of this method to address the connection between cluster morphology and cosmology and to assess whether an individual cluster is sufficiently relaxed for analysis of its intrinsic shape using hydrostatic methods. Approximately 50 X-ray observations of Abell clusters with the PSPC will be amenable to morphological analysis using the method of this paper.

*Subject headings:* galaxies: clustering — galaxies: evolution — galaxies: structure — X-rays: galaxies

---

[1]dbuote@space.mit.edu

[2]jcht@cloud9.arc.nasa.gov

## 1. Introduction

Clusters of galaxies, the largest bound objects in the universe, are especially useful laboratories for probing the underlying cosmology (for reviews see Oegerle, Fitchett, & Danly 1990; Durret et al. 1995). The dynamical state of clusters in particular has received much recent attention because of its implications for estimations of the detailed mass distributions (e.g., Fitchett 1990) and intrinsic shapes (Buote & Tsai 1995a) of clusters, and it may provide a powerful constraint on the cosmological density parameter $\Omega$ (Richstone, Loeb, & Turner 1992, hereafter RLT; Evrard et al. 1993). These studies emphasize that the dynamical state of a cluster is qualitatively related to the degree of substructure present. Hence there is a need for a scheme that quantifies the morphology of clusters in relation to how much "dynamically relevant" substructure they possess.

What constitutes "dynamically relevant" substructure is at present poorly defined. For cosmological purposes RLT make the provisional suggestion that the relevant substructure must have a density contrast compared to the mean cluster density in the range 2-10, contain at least 20% of the cluster mass, and lie within a projected radius of 1.5 Mpc. Denser structures would also be important for studies of the underlying mass distribution and intrinsic shape.

Present techniques to measure substructure in clusters generally provide estimates of the statistical significance of the existence of substructure. The KMM algorithm (Ashman, Bird, & Zepf 1994), for example, detects and quantifies the significance that a given cluster is bimodal. Mohr, Fabricant, & Geller (1993) introduced the centroid shift to quantify substructure in X-ray images of clusters. Although useful for establishing the existence of substructure, these methods do not specifically address the relation of the structure to the dynamical state of the cluster.

We propose a method to quantitatively classify clusters of different morphologies in direct relation to the dynamical state of the cluster as indicated by the gravitational potential. We focus on structure that is obvious in projection; e.g., the morphological classes described by Jones & Forman (1992) for X-ray clusters. That is, the significance of the substructure is a given, what the structure implies for the cluster dynamics is the focus of this paper. The method yields dimensionless quantities that are especially suited to statistical analysis of a large cluster sample. Future papers will apply this technique to a large sample of X-ray clusters (Buote & Tsai 1995b) and to clusters generated by N-body / hydrodynamic simulations. In §2. we describe the method. We demonstrate the performance of the method on simple models of X-ray clusters in §3. and on simulated observations of these models with the ROSAT PSPC in §4.. We apply the method to real ROSAT PSPC images of four Abell clusters in §5.. In §6. we discuss the implications of our method. In §7. we present our conclusions.

## 2. Method

The objective of this paper is not to introduce another technique to detect subtle manifestations of substructure in galaxy clusters (e.g., Ashman et al. 1994; Bird & Beers 1993; for a comprehensive review see Bird 1993), but instead is to propose a simple quantitative scheme to categorize clusters (in projection) of different morphological types; e.g., a smooth single-component cluster or a widely separated bimodal in the plane of the sky. We focus on subclustering that is easily discernible in projection and dynamically important to the whole cluster; i.e. hidden substructure along the line of sight is not our concern nor are individual galaxies and small groups that do not significantly contribute to the potential energy of the cluster. More precisely, we will consider a cluster having two or more components that constitute a sizeable fraction ($\gtrsim 10\%$) of the total cluster gravitational potential within $\sim 1$ Mpc of the cluster center (in projection) to possess the type of substructure relevant to this paper; these criteria should include much of the substructure suggested by RLT as relevant for cosmological studies. We define a cluster possessing this type of substructure to be a *multicluster*.

The natural basis for classifying multiclusters is the two-dimensional multipole expansion of the projected gravitational potential. Let $\Sigma(R, \phi)$ be the two-dimensional projection of the multicluster mass density, where $(R, \phi)$ are the conventional polar coordinates. This $\Sigma(R, \phi)$ generates the two-dimensional potential $\Psi(R, \phi)$,

$$\nabla^2 \Psi(R, \phi) = 4\pi G \Sigma(R, \phi), \tag{1}$$

where $\nabla^2$ is the two-dimensional Laplacian and $G$ is the gravitational constant. The standard analysis using Green's functions (e.g., Jackson 1975) shows that the potential due to material interior to $R$ is,

$$\Psi(R, \phi) = -2G a_0 \ln\left(\frac{1}{R}\right) - 2G \sum_{m=1}^{\infty} \frac{1}{m R^m} \left(a_m \cos m\phi + b_m \sin m\phi\right), \tag{2}$$

where the moments $a_m$ and $b_m$ are given by,

$$\begin{aligned}
a_m(R) &= \int_{R' \leq R} \Sigma(\vec{x}') \, (R')^m \cos m\phi' \, d^2x', \\
b_m(R) &= \int_{R' \leq R} \Sigma(\vec{x}') \, (R')^m \sin m\phi' \, d^2x',
\end{aligned}$$

and $\vec{x}' = (R', \phi')$. This expansion, similar to its cousin in three-dimensions (e.g., Binney & Tremaine 1987), has the general properties that (1) the circularly symmetric monopole term (i.e. the logarithmic term in eq. 2) is always at least as important as higher-order terms, and (2) the dipole ($m = 1$) vanishes when the origin of the coordinate system is set to the center of projected mass. $\Psi$ does not represent the total gravitational potential due to $\Sigma$ since eq. (2) neglects the mass exterior to $R$. However, as indicated below, we will only be concerned with the gravitational effects due to the interior mass.

A single elliptical cluster only contributes to even terms in the multipole expansion for $\Psi$ (eq. [2]) when the origin is defined to be the center of projected cluster mass [3]. Hence, a significant contribution to odd multipole terms unambiguously reflects substructure (asymmetry) in the projected cluster, although a multicluster need not have odd multipoles (e.g., a bimodal cluster composed of equal-sized subclusters). The even multipoles are also important since multiclusters of different morphologies differ in their relative contributions to the even (and odd) multipole terms. In keeping with the above definition of a multicluster, we will only consider the first few multipole moments ($m = 0, 1, 2, 3, 4$) since higher-order terms reflect smaller-scale, dynamically less significant structures (see §3.).

Each term of the multipole expansion (eq. [2]) is a function of position $(R, \phi)$. Since we want to characterize the dynamical state of a large region of a multicluster, a simple procedure is to evaluate the multipole moments in a circular aperture; a circular aperture does not introduce biases and systematic effects inherent when the aperture shape is modified (through iteration) to conform to the shape of an individual multicluster. By computing the moments in a circular aperture of radius $R_{ap}$, the multipole expansion is sensitive to structures having a scale $\lesssim R_{ap}$ and is most sensitive to scales $\sim R_{ap}$; e.g., if $R_{ap}$ is much greater than any scale associated with the cluster then the only significant term in the multipole expansion will be that corresponding to $m = 0$. Since we are uninterested in structure on scales greater than the aperture size, it is sensible to neglect the contribution to the potential of mass exterior to $R_{ap}$ as we have done in the multipole expansion of eq. (2).

Because the $m \geq 1$ terms of eq. (2) vanish when integrated over $\phi$, we instead consider the magnitude of the terms of each order integrated over $\phi$. Let $\Psi_m$ equal the $m$th term in the multipole expansion of $\Psi$. Define the quantity

$$P_{m,m'}(R_{ap}) = \frac{1}{2\pi} \int_0^{2\pi} \Psi_{m'}(R_{ap}, \phi) \Psi_m(R_{ap}, \phi) \, d\phi. \tag{3}$$

Only terms for which $m = m'$ are non-vanishing. Therefore $P_m \equiv P_{m,m}$ measures the "power" within $R_{ap}$ of the terms of order $m$. Ignoring factors of $2G$, these are given by,

$$P_0 = [a_0 \ln(R_{ap})]^2, \tag{4}$$

for $m = 0$ and

$$P_m = \frac{1}{2m^2 R_{ap}^{2m}} \left(a_m^2 + b_m^2\right) \tag{5}$$

for $m > 0$. The total power of all multipole moments is simply given by

$$P = \sum_{m=0}^{\infty} P_m. \tag{6}$$

---

[3] We will, however, make some use of the dipole term (see the following sections).

The ratio $P_m/P$ reflects the contribution of the $m$th multipole moment to the power of the total gravitational potential within $R_{ap}$. Since the gravitational potential is directly related to the dynamical state of a cluster, the $P_m$ are precisely the measures for substructure we seek. Each $P_m$ has units $(mass)^2$ and the scale is set by $R_{ap}$. Since we are primarily interested in the dimensionless ratios of $P_m$ (see §3.) the only dimension that needs to be specified is the scale $R_{ap}$.

Because we ignore structures outside of the aperture of radius $R_{ap}$ we naturally may obtain different indications for the dynamical state of a cluster depending on the aperture size. For example, a widely separated bimodal of equal-sized components may appear essentially relaxed if the aperture is small enough so that it encloses only one of the subclusters. If the aperture is large enough to include both components, then the $P_m$ will show that the cluster is unrelaxed (actually $P_m/P_0$ – see §§3. and 5.). The ability to quantify the dynamical state of the cluster on varying scales is a great asset of the method. If a cluster is virialized then its dynamical state is trivially well defined; i.e. the cluster is relaxed on scales larger than the constituent galaxies. However, for clusters with significant substructure (e.g., the above bimodal with individually virialized subclusters) the dynamical state is a meaningful concept only when referred to a particular scale; the relevant scale for the above bimodal is determined by the relative separation of the subclusters. The cluster on scales smaller than this, or significantly greater than this, may well be virialized.

If we know the projected cluster mass density $\Sigma$, then the physical interpretation of the $P_m$ is manifest. One may non-parametrically construct an accurate map of $\Sigma$ by analyzing the weak distortions of background galaxies (e.g., Kaiser & Squires 1993). Unfortunately, since this technique is in its infancy and the measurement is difficult (requires sub-arcsecond seeing), such maps of the projected cluster mass density exist for only a few clusters (e.g., Fahlman et al 1994; Smail et al. 1995). In any event, the required source-lens-observer distances ($0.15 \lesssim z \lesssim 0.6$, where $z$ is the redshift) fundamentally limits the number of clusters for which weak-lensing maps of $\Sigma$ may be obtained.

In order to analyze the structures of a large sample of clusters we must appeal to either X-ray images or galaxy positions as a practical substitute for $\Sigma$. X-ray maps of clusters have several advantages over galaxy positions. First, the statistical uncertainties associated with galaxy positions are intrinsically fixed by the finite number of cluster galaxies whereas the noise of X-ray images is limited only by the sensitivity of the detector and the exposure time of the observation. Second, the projection along the line of sight of small groups of galaxies not associated with the cluster are less important for X-ray images because the ratio of X-ray luminosity of a cluster to a group is much greater than the corresponding ratio of projected galaxy number densities. Third, in a similar manner, the projection of small groups within the cluster itself is less significant in the X-rays than in the galaxy counts; i.e. the large clumps that are dynamically important to the cluster as a whole are proportionally more important in the X-rays. Although X-ray images are more suitable for our categorization of multiclusters than galaxy positions, neither exactly represents the projected mass density of the multicluster. As a result, we must revise our interpretation of the multipole expansion (eq. [2]) when analyzing X-ray images.

The X-ray surface brightness, $\Sigma_x$, physically differs from $\Sigma$ of clusters in several respects. Since the X-ray gas emissivity is proportional to $\rho_{gas}^2$, small increases in the gas density enhance the X-ray emission correspondingly more than in the underlying mass density. Moreover, the relationship of the gas distribution to the underlying mass is unclear because the gas may trace the underlying matter, the underlying potential, or neither (e.g., Buote & Tsai 1995a). In any event, the gas is unlikely to dominate the cluster gravitational potential.

To achieve successful quantitative classification of multiclusters we require that multiclusters of different morphological types as determined from analysis of the projected mass density ($\Sigma$) be categorized into corresponding morphological types when using X-ray images. However, the actual values of the $P_m$ derived from the X-ray images need not be the same as those derived from $\Sigma$. In order to investigate this issue let us consider the projected gas emissivity, $\Sigma_x$, as the source term of a hypothetical two-dimensional X-ray emissivity potential $\Psi_x$. Consider a single elliptical cluster and a widely separated bimodal multicluster with different sized components. The qualitative features of the multipole expansion for $\Psi_x$ and $\Psi$ are the same: the odd multipole terms will vanish for the single cluster but not for the bimodal multicluster and the relative proportions of the $P_m$ will differ for both clusters. Therefore, if the X-ray emission for a given multicluster exhibits the same qualitative structure present in the projected mass density $\Sigma$, then the multipole analysis of X-ray images should enable quantitative classification of multiclusters in the same manner as multipole analysis of $\Sigma$. We explore the feasibility of this scheme in the following sections.

## 3. Models

We investigate the capacity of multipole decomposition of X-ray images of galaxy clusters to differentiate multiclusters from single-component clusters; by "single-component" we mean one cluster component dominates the gravitational potential in projection. To this end we examine simple models that capture the features of real X-ray clusters essential for evaluating the utility of multipole analysis. Here we shall only consider the fundamental ability of multipole analysis to distinguish the different models. In the next section we address the expected real performance of the technique by analyzing simulated observations of these models using the instrument parameters of the ROSAT PSPC.

X-ray images of clusters exhibit a variety of morphologies (see Forman & Jones 1990; Jones & Forman 1992) ranging from smooth, single-component clusters with regular, nearly elliptical isophotes to multi-component clusters possessing several independent emission peaks. We will focus our attention on the simplest multicluster, the bimodal. A multicluster with more than two components typically has more power in higher order moments and thus is easier to distinguish from a single-component cluster; we will, however, investigate real clusters spanning the observed range of X-ray morphologies in §5.. Therefore, by examining bimodal multiclusters we will obtain a conservative estimate of the viability of multipole analysis.

We employ simple models of the aggregate structure of X-ray clusters. For our purposes the radial surface brightness profiles of X-ray clusters are sufficiently well parametrized by the $\beta$-model (e.g., Cavaliere & Fusco-Femiano 1976; Jones & Forman 1984; Sarazin 1986),

$$\Sigma_x(r) \propto \left[1 + \left(\frac{r}{a_x}\right)^2\right]^{-3\beta+1/2}, \qquad (7)$$

where $a_x$ is the core radius and $\beta$ the slope parameter. From analysis of a sample of bright *Einstein* clusters, Jones & Forman found $\beta \sim 0.5 - 1$ and $a_x \sim 50 - 750$ kpc. Jones & Forman noticed that the smooth, single-component clusters with a dominant central galaxy had the smallest core radii in their sample, $a_x \lesssim 300$. The clusters with large core radii probably possess significant core substructure like Abell 2256 (see Jones & Forman 1991; Briel et al. 1991). As a result, we assign parameters consistent with the smooth single-component clusters having a centrally dominant galaxy to each subclump of a bimodal multicluster. In order to incorporate models having a constant ellipticity ($\epsilon$) and orientation, we substitute for $r$ the elliptical radius $a$, where $a^2 = x^2 + y^2/q^2$, where $q$ is the constant axial ratio.

We compute $P_m$ on a large sample of cluster models. The single-component clusters are represented by single $\beta$-models with $\epsilon = 0.1 - 0.6$, $a_x = 100 - 700$ kpc, and $\beta = 0.5 - 1$. The ellipticity range spans plausible values for a single-component, ellipsoidal, non-rotating, self-gravitating mass (Merritt & Stiavelli 1990; Merritt & Hernquist 1991) appropriate to the case where the gas traces the underlying mass. If the X-ray gas is in hydrostatic equilibrium, then it will trace the potential and thus be rounder $\epsilon = 0 \sim 0.3$; for a discussion of the evolution of X-ray gas shapes see Buote & Tsai (1995a). Although large $a_x$ and $\epsilon$ probably characterize a multicluster, we include these values for single-component clusters so as to make the distinction between multiclusters and single-component clusters more difficult to observe; in this manner we may obtain a conservative evaluation of the method. For the bimodals we consider for each component $a_x = 100 - 300$ kpc and for convenience set each $\beta$ to 0.75. The relative separation of the components ranges from 250 kpc to 1.5 Mpc and their relative normalization ranges from 1:1 to 1:100. We also allowed the primary component in the bimodal to have $\epsilon = 0 - 0.4$ while the secondary component is always kept circular and placed on the major axis of the primary component; note all models were normalized to have the same value of $a_0$ within an aperture of radius 2 Mpc. The range of models we have selected should bracket most bimodals that would be considered multiclusters.

For each model we compute the powers, $P_m$ (eq. [4] and [5]), for $m = 0 - 6$ within a circular aperture of radius $R_{ap} = 1$ Mpc; we also examine $R_{ap} = 2$ Mpc. The projected gravitational potential given by equation (2) is defined up to a constant which we specify by choosing units for $R_{ap}$; note the choice of units for $R_{ap}$ is irrelevant for the $m \geq 1$ terms. In order to ensure that the $P_m$ have the same values for a cluster independent of its distance we express $R_{ap}$ in units of kpc. In Figure 1 we display contours of four models placed at $z = 0.10$ (see §4.).

Each choice for the location of the aperture center gives rise to a different multipole expansion

(eq. [2]). First, we place the origin of the aperture at the centroid; i.e. where $P_1$ vanishes. By doing so, any information possibly found in the first moment is transferred to the higher order terms. The quadrupole power, $P_2$, in this case, is related to the ellipticity of the cluster. The next moment, $P_3$, is sensitive to bimodal structure where the components are of unequal size. $P_4$ is similar to $P_2$ but is more sensitive to smaller-scale structure. Hence $P_2$, $P_3$, and $P_4$ yield complementary information on the structure of the cluster.

We investigate a second case where we place the center of the aperture at the peak of the surface brightness. To avoid confusion with the previous case, moments generated with the origin at the emission peak are denoted $P_m^{(pk)}$. Here, the first non–symmetric moment $P_1^{(pk)}$ only vanishes if the cluster exhibits reflection symmetry about two orthogonal axes centered on the origin (such as for a pure elliptical cluster). This moment is then particularly sensitive to bimodal multiclusters having nearly equal-sized components. $P_1^{(pk)}$ essentially characterizes a circularly-averaged, centroid-shifting power of the cluster within $R_{ap}$. Although there exists some combination of the $P_m$ (centroided) that contains the information given by $P_1^{(pk)}$, we employ $P_1^{(pk)}$ because of its higher sensitivity to nearly equal-sized bimodal multiclusters relative to $P_2$, $P_3$, and $P_4$.

We list in Table 1 the powers $P_2$, $P_3$, and $P_4$ computed for a selection of the above models expressed as a ratio of $P_0$. These values are indicated by "True" in the table. In Table 2 we list the power $P_1^{(pk)}$ in terms of $P_0^{(pk)}$, where $P_0^{(pk)}$ is the $m = 0$ power computed in the circle of radius $R_{ap}$ centered at the emission peak. We prefer to consider power ratios, $P_m/P_0$, instead of the individual $P_m$ because (1) dividing by $P_0$ normalizes to the flux within $R_{ap}$ which enables consistent comparison between clusters of different X-ray brightnesses, and (2) the ratio more easily distinguishes between an image that is centrally concentrated (i.e. lower $P_m/P_0$) to one that is more spread out (i.e. higher $P_m/P_0$); e.g., even if a single and bimodal cluster (of the same luminosity) have the same $P_2$ within $R_{ap}$, the bimodal will necessarily have a smaller $P_0$. Moreover, clusters that are more relaxed should be more dominated by their monopole terms. We do not list results for the higher order moments ($m \geq 5$) since they are, as expected, particularly sensitive to the bimodal multiclusters having relative normalizations greater than 10 : 1. These models are inconsistent with our definition of a multicluster or the substructure described by RLT and are also more susceptible to the hazards of real data (e.g., noise, unresolved sources – see next section).

As expected the odd terms yield the largest differences between single and bimodal clusters since they vanish identically for the single-component clusters. The results also demonstrate the complementary behavior of $P_1^{(pk)}/P_0^{(pk)}$ and $P_3/P_0$; i.e. the former is larger for equal-sized bimodals while the latter vanishes; the former is much less sensitive for unequal-sized bimodals while the latter is most sensitive to them. The even ratios, $P_2/P_0$ and $P_4/P_0$, also perform well at distinguishing bimodals from low-ellipticity ($\epsilon \lesssim 0.3$) single-component clusters. However, there is considerable overlap of the flatter single-component clusters ($\epsilon \gtrsim 0.3$) and moderately separated bimodals (0.5 Mpc separation of each subclump). For 1 Mpc separation the degeneracy is lifted and the even ratios separate well the single-component and bimodal models.

From the results listed in Tables 1 and 2, it is clear that the power ratios ($m \leq 4$) easily discriminate bimodal multiclusters from single-component clusters; here we considered multiclusters to have relative separations $500 < r < 1500$ kpc and relative normalizations $< 10:1$. Moreover, they do not significantly distinguish bimodals with relative separations $< 500$ kpc or relative normalizations $> 10:1$ from single-component clusters implying that these power ratios are essentially only sensitive to structure relevant to multiclusters; although we mention that $P_3/P_0$ and $P_4/P_0$ are sensitive to some models with larger relative normalizations when the components are sufficiently well separated.

The ability of the power ratios to distinguish models is dependent on the aperture size. In Table 3 we list the power ratios as a function of aperture size (radii 0.5,1.0,1.5, and 2.0 Mpc) for two models of very different morphologies: a single-component cluster with $\epsilon = 0.3$, core radius 300 kpc, and $\beta = 0.75$; and a bimodal where both components have $\epsilon = 0$, core radius 300 kpc, $\beta = 0.75$, are separated by 1 Mpc, and have relative normalization $2:1$. The power ratios of the single-component model (which naturally has vanishing odd $P_m$) are largest for the smallest apertures and decrease monotonically with increasing aperture size. This behavior simply reflects the decay of higher-order multipole terms with distance just as if the X-rays represented a true mass distribution (see §2.); i.e. the qualitative picture of the multipole expansion is preserved for our simple models. In the 0.5 Mpc aperture, the bimodal cluster has smaller values of $P_2/P_0$ and $P_4/P_0$ than the single-component cluster because the aperture only encloses the dominant subcluster; i.e. these power ratios are similar to low-ellipticity single-component models (see Tables 1 and 2). However, the small non-zero values of the odd power ratios, $P_1^{(pk)}/P_0^{(pk)}$ and $P_3/P_0$, demonstrate that some indication of a large-scale asymmetry is detected within the 0.5 Mpc aperture. At 1 Mpc the centroided power ratios peak and then decay for larger apertures illustrating the power-ratios' sensitivity to the scale of the substructure (i.e. 1 Mpc separation of the two components). As expected, the higher-order moments decay most rapidly. In contrast, $P_1^{(pk)}/P_0^{(pk)}$ peaks at $\sim 1.5$ Mpc but otherwise behaves similarly to the centroided moments.

## 4. Simulated Observations

Although we have demonstrated the ability in principle of the power-ratio method to quantitatively differentiate single-component clusters from multiclusters, we have not shown if these statistics can by usefully constrained by present observations. We now assess the practical feasibility of the power-ratio method by simulating real X-ray observations of the cluster models with the ROSAT Position Sensitive Proportional Counter (PSPC); for a description of ROSAT see Trümper (1983), of the ROSAT X-ray telescope see Aschenbach (1988), and of the PSPC see Pfeffermann et al. (1987). The PSPC is more suited to this kind of study than the *Einstein* Imaging Proportional Counter because of its superior resolution ($\sim 30''$ FWHM, on axis) and sensitivity. The sensitivity of the PSPC even outweighs the better spatial resolution ($\sim 4''$ FWHM) of the ROSAT High Resolution Imager (HRI). Moreover, many more clusters have been observed with the PSPC (see §6.) than the HRI. The PSPC is also well suited for our study because the energy band pass and spectral response implies that the observed $\Sigma_x$ depends almost exclusively on $\rho^2$ and is independent of temperature. Hence variations in $\Sigma_x$ will not be due to temperature fluctuations in the gas.

Since our primary motive for introducing the power-ratio method is to facilitate consistent comparison of the morphologies of statistical samples of clusters (see §1. and §6.), we simulated observations of a well-defined cluster sample. We also chose our observational parameters to take advantage of the large number of cluster observations that are becoming available in the ROSAT archive. In keeping with the spirit of the power-ratio method (see beginning of §2.), we concentrated on bright (for $S/N$) and nearby ($z \lesssim 0.2$; for resolution) clusters. We considered clusters having characteristic X-ray luminosity $L_x^\star$ obtained from fitting the Schechter luminosity function to Abell and ACO clusters in the ROSAT All-Sky Survey (RASS; Ebeling 1993). Ebeling obtains $L_x^\star = 1.69 \pm 0.37 \times 10^{44}$ erg s$^{-1}$ for his sample which translates to a flux $F_x^\star = 1.6 z^{-2} \times 10^{-13}$ erg cm$^{-2}$ s$^{-1}$, where $z$ is the cluster redshift and we have assumed a linear distance-redshift relation; this assumption amounts to assuming a static Euclidean universe which is suitable for our purposes to explore the effects of angular scale and $S/N$ on the power ratios of low-redshift clusters. We placed $L_x^\star$ clusters at a series of redshifts ($z = 0.05, 0.10$, and $0.20$) appropriate for most of the bright clusters that will be available in the ROSAT archive; a Hubble constant $H_0 = 80$ km s$^{-1}$ Mpc$^{-1}$ is assumed so these redshifts correspond to 187.5, 375 and 750 Mpc respectively. To agree with typical observations, we simulated observations having exposure times ($t_{exp}$) of 5ks, 10ks, and 20ks.

We constructed a simulated observation for a given model by first choosing the redshift of the cluster and then the exposure time. The flux was converted to PSPC counts in the hard band (0.4 - 2.4 keV) using the energy-conversion factors ($ECFs$) given in NRA 91-OSSA-3, Appendix F, ROSAT mission description. For each cluster we assumed a thermal line spectrum for temperatures $T = 4 - 8$ keV and column densities $N_H = 10^{19.5} - 10^{21}$ cm$^{-2}$ which translates to a typical $ECF = 0.575 \times 10^{11}$ counts cm$^2$ erg$^{-1}$; the total counts were then $(F_x^\star)(ECF)(t_{exp})$. The counts within 1 Mpc of the centroid were normalized to the above total counts. We added a uniform background with a count rate of $3 \times 10^{-4}$ counts s$^{-1}$ arcmin$^{-2}$ which is the average of the background rates of two PSPC images previously studied (Buote & Canizares 1994,1995). The models were then convolved with the off-axis point spread function of the PSPC evaluated at 1 keV (MPE/OGIP Calibration Memo CAL/ROS/93-015); note we did not include the support structure of the PSPC and thus we confined our study to within the inner $40'$ diameter ring of the PSPC. Moreover, we did not add any exposure variations or vignetting since we assumed the observer can adequately correct for these effects. We set the pixel scale to $15''$ for all the models.

We included the effects of noise from point sources and Poisson statistics in the simulations. First, to each image we added point sources having spatial properties consistent with the PSF of the PSPC and numbers consistent with the $\log N(>S) - \log S$ distribution given by Hasinger (1991) from analysis of the RASS; see Soltan & Fabricant (1990) and Mohr, Fabricant, & Geller (1993) for the inclusion of point sources in simulated *Einstein* images. The flux of each source was determined randomly from the probability distribution $dN(>S)/dS$ and then positioned randomly in the field. As before, we converted fluxes to PSPC counts using the $ECF$s in NRA 91-OSSA-3. We modeled the point sources with power-law spectra having indices $1.1 - 1.9$ and $N_H = 10^{19.5} - 10^{21}$ cm$^{-2}$ which gave $ECF$s ranging from $(0.40 - 0.55) \times 10^{11}$ counts cm$^2$ erg$^{-1}$. For each point source we randomly selected an $ECF$ from these values. Any point source that would be detected by an observer we excluded from the image, although we did count such a source toward the total number of sources dictated by $N(>S)$. We evaluated the significance of each source by comparing the source counts to the total counts in a circle of radius $30''$ centered on the source. We conservatively excluded any source $\geq 5\sigma$ above the total noise.

For each model corresponding to a particular redshift and exposure time we generated 1000 simulated observations, and for each simulated observation we computed the power ratios $P_1^{(pk)}/P_0^{(pk)}$, $P_2/P_0$, $P_3/P_0$, and $P_4/P_0$ as described in §3.; here we set $R_{ap} = 1$ Mpc. Although a uniform background will not contribute to $P_m$ for $m \geq 1$, it does contribute to $P_0$. Hence, we subtracted the mean background to ensure proper normalization of the power ratios. To evaluate $P_2/P_0$, $P_3/P_0$, and $P_4/P_0$ we computed the centroid within $R_{ap}$ and iterated until the centroid shifted by less than $10^{-5}$ pixels (or 25 iterations had been performed).

We were compelled to adopt a more sophisticated procedure for $P_1^{(pk)}/P_0^{(pk)}$ because the position of the emission peak is much more sensitive to noise then the centroid. For determination of the peak position we first convolved the simulated image with a Gaussian filter having $\sigma$ corresponding to a physical dimension of 40 kpc. (For example, this corresponds to 3 pixels for $z = 0.05$.) Then we took the highest 25% of the pixels within a circle of radius $0.15\,R_{ap}$ centered at the peak of the smoothed image and computed the centroid of the highest points. We defined this centroid to be the emission peak. Using this position for the peak, we computed $P_1^{(pk)}/P_0^{(pk)}$ on the original unsmoothed simulated image.

We computed simulated observations of a subset of models from §3. possessing the essential distinguishing characteristics of the single-component and multicluster models. The single-component models have $\epsilon = 0, 0.3$, and $0.6$ while the core radius was set to 300 kpc and $\beta$ to 0.75. The multiclusters have for both cluster components $\epsilon = 0$, core radii set to 300 kpc, $\beta$ set to 0.75, relative separations 0.5 and 1 Mpc, and relative normalizations 1:1, 2:1, and 5:1. In Figure 2 we display contour plots of simulated observations for four of these cluster models.

The mean value and the 90% confidence limits of the power ratios computed for the 1000 simulated observations of each model are listed in Tables 1 and 2; for lack of space we only include the $t_{exp} = 10$ks exposures. The 90% confidence limits are specified so that 10% of the simulated observations give power ratios above the upper limit and 10% give power ratios that are below. It is clear that for our chosen sample of bright, nearby clusters the power ratios perform nearly as well on the simulated observations as on the exact models (§3.) for distinguishing multiclusters from single-component clusters. In fact, the mean value of $P_2/P_0$ is generally within $\sim 10\%$ of the true values and the 90% confidence limits bracketed the true value in all but a few cases. $P_1^{(pk)}/P_0^{(pk)}$, $P_3/P_0$, and $P_4/P_0$ behave similarly, although they are usually only within $\sim 25\%$ of the true values. In addition, for $P_1^{(pk)}/P_0^{(pk)}$ and $P_3/P_0$ the observations give noticeable non-zero values (as high as a few times $10^{-7}$ for $P_3/P_0$ and a few times $10^{-6}$ for $P_1^{(pk)}/P_0^{(pk)}$ ) for some models where the true value is identically zero. The point sources and Poisson noise do indeed contribute noticeable uncertainty, but the uncertainties are generally less than the systematic differences in power ratios between the two classes of models. We mention that the contribution to noise in the power ratios is roughly equal between the point sources and Poisson noise. We also find that the effects of using pixelized images to determine power ratios are very small for the chosen redshifts. We do this by essentially determining the difference between a locally smoothed version of the image to the observed pixelized image.

The results for $P_1^{(pk)}/P_0^{(pk)}$ listed in Table 2 demonstrate that the systematic effects due to locating the emission peak are insignificant with respect to noise, pixelization, etc.. This systematic effect is due to noise and pixelization of the image only in the sense that the emission peak is difficult to determine accurately from a realistic observation. That is, our procedure for determining the center by smoothing and then centroiding on the highest surface brightness pixels tends to move the center closer to high surface brightness structures which are near the absolute peak. For example, if there are two high emission structures which are not extremely widely separated, say in a bimodal cluster, then the center will be moved in the direction of the centroid of the image. This leads to a smaller amount of power in the given ratio. This effect does not influence cases such as the single elliptical cluster where the center is well separated from other emission peaks. We emphasize that given realistic observations, the ratio $P_1^{(pk)}/P_0^{(pk)}$ is a good discriminator between single clusters and bimodals, especially those bimodals with nearly equal sized components, and brackets the results of the true models within the $\sim 90\%$ confidence level.

We examine whether the sample of multiclusters may be distinguished from the sample of single-component clusters by applying a Kolmogorov-Smirnov test to the individual distributions of power ratios. That is, we have 1000 simulated observations of 27 single-component cluster models and 54 bimodal multicluster models which form two well-defined cluster samples that differ only in their morphology, not in their luminosity or distribution or distances; the bimodal models with relative normalizations 2:1 were omitted for $P_1^{(pk)}/P_0^{(pk)}$ leaving 36 total bimodals for that case. We obtained, for example, the distribution of $P_3/P_0$ values for each sample by selecting $P_3/P_0$ randomly from one of the 1000 simulations of each model. We list in Figure 3 the cumulative distributions of $P_1^{(pk)}/P_0^{(pk)}$, $P_2/P_0$, $P_3/P_0$, and $P_4/P_0$ obtained in this manner The K-S tests yield probabilities of $< 0.001\%$ for $P_1^{(pk)}/P_0^{(pk)}$, 3% for $P_2/P_0$, 0.03% for $P_3/P_0$, and 0.1% for $P_4/P_0$ that the singles and bimodals originate from the same population. Thus the K-S test for each power ratio convincingly demonstrate that the two data sets could not have come from the same parent population. Although the K-S test is useful for hypothesis testing, in §6. we discuss the correlations of the power ratios as a means to provide detailed classification of clusters into different morphological types.

## 5. Application to Four Real Clusters

As a final illustration of the performance of the power-ratio technique, we analyzed ROSAT PSPC X-ray images of the four Abell clusters A85, A514, A1750, and A2029. A2029 is a classic single-component cluster having regular, moderately flattened ($\epsilon_x \sim 0.15$) elliptical isophotes; it is nearby ($z = 0.0768$) and very X-ray bright ($\sim \frac{5}{2}L_x^\star$; the fluxes of the clusters are listed in Table 5). The other three clusters represent the quintessential examples from the qualitative morphological classification scheme of Jones & Forman (1992). A85 ($z = 0.0556$) has a dominant primary component with a small secondary and is one of the brightest clusters ($\sim \frac{3}{2}L_x^\star$). A1750 has two components of roughly equal size with a respectable X-ray luminosity ($\sim \frac{2}{3}L_x^\star$). Finally, A514 is a complex X-ray cluster as classified by Jones & Forman because it has at least three distinct emission peaks and highly irregular isophotes; it is at similar redshift as the others ($z = 0.0731$) but is somewhat fainter in X-rays ($\sim \frac{1}{6}L_x^\star$). Therefore, the four clusters span the range of observed X-ray morphologies of clusters and hence serve as convenient benchmarks for demonstrating the ability of the power-ratios to quantitatively classify real clusters.

We prepared the images for analysis using the standard IRAF-PROS software. First we searched for spikes in the light curves of the images that signal contamination from solar radiation; no statistically significant spikes were found for the images. Next we rebinned the images into $15''$ pixels corresponding to $512 \times 512$ fields; i.e the same scale used for the simulated observations in §4.. Only data from the hard band ($0.5 - 2$ keV) were used in order to minimize the blurring due to the point spread function (PSF) of the PSPC and the contamination from the X-ray background. We corrected for exposure variations and telescopic vignetting by dividing the images by the exposure maps provided with the observations. To subtract the background we selected an area $\sim 35'$ from the field centers apparently uncontaminated by emission from nearby sources or the cluster itself. Typically we identified the background region from examination of the radial profile centered on the cluster; we designated the background region where the radial profile flattened. This procedure undoubtedly suffers to some extent from contamination by emission from the cluster, but the errors are insignificant in relation to the total cluster flux which is all that we require (see §4.). In Table 4 we list the observational data for the four clusters; in Figure 4 we show contour plots of the images.

The final step in the image reduction is to remove embedded sources from the cluster continuum. In keeping with the simulated observations of the previous section, we identified and removed "obvious" contaminating point sources from visual examination of the images; note we arrived at this subjective procedure because automated techniques like the *detect* package in PROS had difficulty identifying sources located in the continuum of the clusters. Although some of these sources that we remove are either noise or intrinsic features of the clusters, removing them only serves to smooth out the cluster emission thereby decreasing the power from higher-order moments; i.e. these small effects only make a multicluster appear more similar to a single-component cluster. Hence, the quantitative significance of a multicluster as separate from a single-component cluster is made more robust. The uncertainties and biases regarding source removal are investigated in detail in our companion paper (Buote & Tsai 1995b). In Figure 5 we show contour plots of the four clusters with the embedded sources removed.

In Table 5 we list the power ratios computed on these four images for $R_{ap}$ of 0.5, 1, and 1.5 Mpc, where we have used $H_0 = 80$ km s$^{-1}$ Mpc$^{-1}$ as in the previous section. From examination of the simulations in §4. we estimate that $\sim 50\%$ uncertainties should reflect the 90% confidence limits for the clusters. The clusters are clearly differentiated by their power ratios. For example, A514 and A2029 have $P_1^{(pk)}/P_0^{(pk)}$ and $P_3/P_0$ values that generally differ by over two orders of magnitude and the $P_2/P_0$ values differ by one order of magnitude, a discrepancy that is highly significant considering the $\sim 50\%$ uncertainties. In fact, $P_1^{(pk)}/P_0^{(pk)}$, $P_2/P_0$, and $P_3/P_0$ easily distinguish A2029 from the each of the other clusters in each aperture. In the 1 Mpc aperture $P_4/P_0$ for A514 is one hundred times larger than the value for A2029, although the difference in $P_4/P_0$ is negligible in the other apertures.

A85 is a prime example of the type of cluster described by RLT as being relevant for cosmological tests (§1.). Moreover, next to A2029, it is the cluster with the least obvious structure in our sample. In the 0.5 Mpc aperture the $P_2/P_0$ and $P_4/P_0$ values for A85 are completely consistent with those of the smooth A2029. The other two ratios, however, demonstrate substantial discrepancy with A2029. All of the power ratios easily distinguish A85 from A2029 in the 1 Mpc aperture with the ratio $P_4/P_0$ being in marginal agreement. Hence the power ratios clearly distinguish A85 from single-component clusters like A2029.

The example of A85 demonstrates that the power ratios are indeed sensitive to the aperture size and classify the clusters according to the scale of the substructure they possess as we demonstrated for the toy models in §3. and Table 3. A more extreme case is A1750 where the 0.5 Mpc aperture only encloses one of the subclusters. The power ratios for A1750 on this scale are completely consistent with those of A2029 except for $P_3/P_0$. It is because a bridge of emission connects the two subclusters that $P_3/P_0$ is able to measure an asymmetry signaling the presence of the subcluster outside the aperture. Unlike A1750, A514 registers very large power ratios in all apertures except for $P_4/P_0$ which is only larger in the 1 Mpc aperture. The sensitivity to the aperture size in this case is a result of the edge of the 1 Mpc aperture falling right on the two subclusters to the West.

The value of $P_1^{(pk)}/P_0^{(pk)}$ for $R_{ap} = 0.5$ Mpc for A2029 in Table 5 appears to significantly differ from the simulations of single-component clusters in the previous section. For several reasons, we do not take this is as an indication of subtle substructure in the central regions of A2029, although such structure is in any case not our primary interest. First, A2029 is known to contain a massive cooling flow (Sarazin, O'Connell, & McNamara 1992) which implies that the surface brightness of the central few hundred kpc does not follow that of the $\beta$ model used in our simulations.

Moreover, Sarazin et al. find complicated structure in the X-ray emission that may be due to any of a number of possibilities (e.g., lumpy absorption or magnetic fields) not intimately related to the total cluster gravitational potential. Since our simulations were restricted to $\beta$ models, they may not have adequately accounted for the variety of surface brightness laws to be found in actual clusters, especially in the central regions. Hence the range of values quoted in Table 2 for the single clusters may not extend to higher values because of the restricted set of models considered and thus the presence of subtle substructure should not be concluded. Despite this slight discrepancy with the simulations, the relatively small values of $P_1^{(pk)}/P_0^{(pk)}$ clearly distinguish A2029 from the other real clusters.

## 6. Discussion

The power-ratio method differs from conventional techniques for analyzing substructure in galaxy clusters because it is designed to quantitatively label clusters of different aggregate morphologies, not simply to quantify the significance of substructure. Moreover, the power-ratio method is motivated by cluster dynamics. If the cluster surface mass density is known (e.g., from weak-lensing maps), the power-ratio method classifies structure in direct proportion to its contribution to the cluster gravitational potential. For X-ray images of clusters this relationship to the gravitational potential is not so clear (§2.), but if the X-rays trace the structure in the surface density then the dynamical interpretation is qualitatively preserved; this was demonstrated by the success of the power ratios at discriminating between the X-ray cluster models and simulations in §3. and §4..

We envisage the power-ratio method to quantitatively distinguish clusters where the substructure is obvious to the eye. For this purpose the significance of the substructure is a given, what this structure implies for the aggregate cluster dynamics is our primary concern. For example, a definitive measurement of a non-zero value of $P_3$ unequivocally demonstrates that substructure is significant for the cluster, but only in relation to the other $P_m$ can it be determined whether the substructure is meaningful to the cluster on a particular scale ($R_{ap}$); the same argument applies to $P_1^{(pk)}$ and therefore to centroid shifts (Mohr et al. 1993). Thus, methods that are particularly suited to locate subtle manifestations of substructure and quantifying its significance (e.g., the KMM algorithm described by Ashman et al. 1994; also see Bird 1993) are not as well suited as the power-ratio method for quantitatively classifying clusters of different morphologies.

To fully realize the capacity of this technique to distinguish clusters by their dynamics, the correlations of the power ratios, rather than each ratio alone, should be analyzed; cf. the K-S tests in §4.. We illustrate these correlations from analysis of the projections of the vector $(P_2/P_0, P_3/P_0, P_4/P_0,)$ onto the two-dimensional coordinate planes for a sample of single-component cluster and multicluster models computed in §3.; we exclude $P_1^{(pk)}/P_0^{(pk)}$ because it necessarily has a trivial correlation with the $P_m/P_0$. The sample of cluster models we now consider is slightly more extensive then shown in Tables 1 and 2 to better illustrate the complete range of cluster behavior. We consider single cluster models with $\epsilon = 0.1, 0.2, 0.3$, $a_x = 100, 300$ kpc, and $\beta = 0.75$; we include only single-component models with $\epsilon < 0.3$ since flatter X-ray clusters probably are not relaxed (e.g., Buote & Tsai 1995a). The bimodals have $a_x = 300$ kpc, $\beta = 0.75$, relative separations $0.5, 1$ Mpc, relative normalizations $1:1, 2:1, 5:1$, and $10:1$, and $\epsilon = 0, 0.2$ for the primary component. For the bimodal models with an elliptical primary ($\epsilon = 0.2$) we consider the cases where the secondary is either aligned along the major axis or the minor axis of the primary. In Table 7 we list the true power ratios for these models computed in apertures of radii $R_{ap} = 1, 2$ Mpc; again we defined $R_{ap}$ in units of kpc.

We plot the correlations of the power ratios in Figure 6 for $R_{ap} = 1$ Mpc; since the singles (represented by filled ovals) vanish for $P_3/P_0$ they have been placed on the $P_2/P_0$ or $P_4/P_0$ axes to show their range in these ratios. In each case the power ratios display positive correlations in the sense that clusters with small relative separations and normalizations inhabit the lower left of the plots while the top right is populated by the nearly equal-sized, widely separated bimodals; i.e. generally the dynamically "mature" clusters are located near the bottom left of the plots while "young" clusters populate the upper right. The tightest correlation is in the $P_2/P_0 - P_4/P_0$ plane for the single-component models where essentially $P_2/P_0 \propto P_4/P_0$. The bimodal multiclusters, although following the same general trend, have much larger scatter especially when the relative normalization $> 5:1$. The correlations clearly show that the multiclusters primarily inhabit a localized region of the three-dimensional space of power ratios; this space is typically $P_2/P_0 = 10^{-6} - 10^{-4}$, $P_3/P_0 = 10^{-8} - 10^{-5}$, and $P_4/P_0 = 10^{-9} - 10^{-6}$. Clusters with smaller relative separations ($\lesssim 0.5$ Mpc) and larger relative normalizations ($\gtrsim 10:1$) generally lie outside this volume in the direction of small values of the power ratios.

In Figure 6 we draw a dashed box to represent the region inhabited by single-component clusters considering the effects of noise, pixelization etc. from simulated observations of these clusters with the PSPC (see §4.). Although some of the multiclusters lie in this region in one of the planes, they are usually removed from correlation with the remaining power ratios. For example, the bimodal with relative separation 1 Mpc, relative normalization $1:1$, and primary $\epsilon = 0.2$ with secondary on the major axis is the model with the lowest $P_3/P_0$ and $P_4/P_0$ shown in Figure 6 and within the error box for single-component clusters in the $P_2/P_0 - P_4/P_0$ plane. However, this model lies outside the error boxes in the other planes and is thus distinguished as a multicluster. In general, where the clusters lie in the volume classifies them by the type of structure they possess; i.e. clusters falling in the multicluster region possess substructure that is dynamically relevant to the aggregate cluster dynamics on a scale $\sim R_{ap}$ (see §2.).

The correlations of the power ratios for $R_{ap} = 2$ Mpc are displayed in Figure 7. Although the power ratios still exhibit positive correlations (with values about 10 times less than before), the distinction between the single-component models and the bimodals is not as pronounced as for the $R_{ap} = 1$ Mpc case – especially upon considering the effects of real observational uncertainty (i.e. as above the dotted lines show the allowed region for single-component clusters considering simulated observations in §4.). This is again a manifestation of the multipole description at work. That is, since the bimodal models we have constructed have separations either 0.5 or 1 Mpc, the higher-order multipoles are largest on those scales. For aperture sizes larger than these scales the monopole term quickly dominates as the higher-order moments rapidly decay. It is thus important to examine different aperture sizes to determine on which scale substructure is particularly important for real clusters. (We mention that the bimodal model with $\epsilon = 0.2$, $R_s = 0.5$ Mpc, and $REL = 2:1$ and secondary aligned along the major axis of the primary has a $P_3/P_0$ value larger at $R_{ap} = 2$ Mpc than $R_{ap} = 1$ Mpc – although both values are very small. This appears to be an interesting case where the ellipticity of the primary acts to reduce the value of $P_3/P_0$ which does not happen for the models where the secondary is aligned along the minor axis of the primary. In a similar case, the bimodal model with $\epsilon = 0.2$, $R_s = 1.0$ Mpc, $REL = 10:1$ and secondary aligned along the minor axis of the primary has $P_2/P_0$ value larger at $R_{ap} = 2$ Mpc than $R_{ap} = 1$ Mpc. More interesting, though, is that this model has a smaller value of $P_2/P_0$ than $P_4/P_0$ for $R_{ap} = 1$ Mpc. We find that these interesting cases are not well-represented by real clusters [see Buote & Tsai 1995b] which may simply be the result of the cluster initially collapsing along its shortest axis – e.g., Lin, Mestel, & Shu 1965.)

A potential application of the power-ratio method is for determining quantitatively whether a quasi-hydrostatic equilibrium description for the X-ray emitting gas of an individual cluster is justified for the purpose of constraining its intrinsic shape and its total mass distribution. Buote & Tsai (1995a) tested the viability of X-ray analysis for constraining the intrinsic shapes of clusters of galaxies using the simulation of Katz & White (1993). They concluded that at low redshifts ($z \lesssim 0.25$) the X-ray method accurately measured the true ellipticity of the three-dimensional cluster dark matter up to projection effects. At higher redshifts ($z \gtrsim 0.25$), however, the X-ray method yielded unreliable results since the gas does not trace the cluster gravitational potential. Buote & Tsai proffer some necessary conditions for the reliability of X-ray methods: (1) that there is no obvious substructure on the same scale used to compute the aggregate shape and (2) the isophotes are regularly shaped and not too elongated ($\epsilon_x \lesssim 0.3$). The power-ratio method is particularly suited to quantify these necessary conditions. Results on our study will be presented elsewhere, but see Buote & Tsai (1995b) for an outline of a prescription for this program.

The power-ratio method is ideally suited to constrain $\Omega$ via the Morphology - Cosmology connection (see §1.; RLT; Evrard et al. 1993). The method provides a simple, consistent comparison of the structure of clusters since the power ratios are (1) computed in a well-defined aperture, (2) normalized to the flux within that aperture, and (3) do not require any fitting. Because it is particularly sensitive to structure relevant to the dynamical state of the cluster, the power-ratio method specifically quantifies the type of substructure described by RLT as relevant for cosmology. The ratio $P_3/P_0$ (and $P_4/P_0$), being sensitive to unequal-sized bimodal multiclusters, is most relevant to the structure envisioned by RLT, while $P_1^{(pk)}/P_0^{(pk)}$ and $P_2/P_0$ are more sensitive to roughly equal-sized subclumps.

A large number of clusters similar to those of our sample in §4. will be available in the ROSAT archive. From examination of the ROSAT master log of pointed observations (in the HEASARC-Legacy database) for Abell clusters having (1) measured flux $\gtrsim 10^{-11}$ erg cm$^{-2}$ s$^{-1}$ as published by Ebeling (1993), (2) exposure times $> 5$ks, and (3) $z \lesssim 0.2$ we find $\sim 50$ eligible clusters. Higher redshift clusters will be available for analysis with AXAF because of its superior resolution. As a result, at least 124 Abell clusters from Ebeling (1993) having flux $> 10^{-11}$ erg cm$^{-2}$ s$^{-1}$ will in principle be eligible for analysis. Thus application of the power-ratio method to these samples should enable a thorough statistical investigation of the viability of using the observed structure of clusters to place interesting constraints on $\Omega$. In Buote & Tsai (1995b) we apply the power-ratio method to 55 ROSAT PSPC clusters to furnish a catalog of power ratios suitable for statistical analysis.

An effort to analyze the Morphology - Cosmology connection was undertaken by Evrard et al. (1993). These authors proposed to quantify the structure of *Einstein* clusters using a mean centroid shift (Mohr et al. 1993), a mean axial ratio, and a mean slope of the surface brightness of the entire X-ray image; the mean centroid shift and mean axial ratio are related to our $a_1^{(pk)}$, $b_1^{(pk)}$ and $a_2$, $b_2$. By computing the power ratios in apertures defined by the cluster distances, we consistently sample the same intrinsic scales of clusters. In addition, by using a series of aperture sizes we also obtain information regarding the scale of the substructure in the cluster sample. Evrard et al., in contrast, sample different cluster scales for each cluster because they compute mean quantities for the entire X-ray images; the size of a cluster X-ray image is dependent on the flux, intrinsic size, and distance of the cluster. Evrard et al. also do not employ a third (or forth) moment which is in fact more sensitive to the unequal-sized bimodal multiclusters envisioned by RLT than the lower-order moments.

The manner in which Evrard et al. compute the centroid-shift and axial ratio (as explained in Mohr et al. 1993) also makes a direct comparison to intrinsic properties of the cluster uncertain. Specifically, the surface brightness in a circular annulus of a given width is first Fourier expanded to low order given a trial center for the annulus. The expansion is then fitted to the image taking the Fourier coefficients as parameters of the fit. The location of the center of the image is then iterated so as to minimize the coefficient of the $m = 1$ term ($C_1$). This center is then used to compute the centroid shift and the axial ratio. This procedure gives correct values for these latter quantities only if the cluster being considered is very nearly elliptical and higher order terms in the Fourier expansion of the surface brightness are small. This is because the values of the coefficients determined by *fitting* a highly truncated version of the expansion to the image are not necessarily the true values of the Fourier coefficients when higher order terms are important. Given a center, the true Fourier coefficients are given by moments of the surface brightness distribution; fitted values will depend on the highest order considered and will not have their usual meanings.

Mohr et al. (1993) only considered clusters for which the condition that higher order terms in the expansion of the surface brightness be small is satisfied. Their values of the centroid shift and the axial ratio are probably accurate. However, consideration of substructure in these clusters are of limited use in cosmological considerations since the substructure required by RLT will indeed give rise to significant higher order terms, as seen in §3.

## 7. Conclusions

We have described a technique to quantitatively classify clusters of galaxies according to their projected morphology. In particular, we addressed structure that is easily discernible in projection and dynamically important to the whole cluster. A cluster possessing substructure of this type (§2.) we defined to be a *multicluster*. We specifically designed our method to quantitatively distinguish multiclusters from single-component clusters in projection. The method is derived from the two-dimensional multipole expansion of the projected cluster gravitational potential; i.e. the square of each multipole term averaged over a circular aperture is called the *power*, $P_m$, and when divided by another $P_m$ (particularly $P_0$) we call it a *power ratio*. For the case where the surface mass density of the cluster is known, e.g., from analyzing the weak distortions of background galaxies (e.g., Kaiser & Squires 1993), the power-ratio method classifies structure in relation to its contribution to the cluster gravitational potential.

For X-ray images of clusters this relationship to the gravitational potential is not so transparent (§2.), but if the X-rays approximately trace the structure in the surface mass density then qualitatively the dynamical interpretation is preserved. We demonstrated this assertion by analyzing the performance of the power-ratio method applied to simple models capturing the essential features of real X-ray clusters. In particular we focused on models of single-component clusters and bimodal multiclusters. By construction the structure of these models of X-ray clusters reflected structure in the projected mass; clumps in the X-rays corresponded to the same clumps in the mass, although not in exactly the same proportions. We determined that the ratios $P_1^{(pk)}/P_0^{(pk)}$, $P_2/P_0$, $P_3/P_0$, and $P_4/P_0$ performed best for distinguishing between single-component clusters and multiclusters; the powers were computed in a circle located at the cluster centroid except for $P_1^{(pk)}/P_0^{(pk)}$ which was centered at the emission peak. The ability of the power ratios to differentiate clusters was optimized when the aperture size was of order the separation of the clumps of the bimodals.

We simulated observations of these models using the instrument parameters of the ROSAT PSPC for a sample of nearby ($z \leq 0.2$) and bright (flux = $F_x^\star = 1.6z^{-2} \times 10^{-13}$ erg cm$^{-2}$ s$^{-1}$) clusters. The effects of point sources, X-ray background, Poisson noise, and realistic exposure times were also incorporated into the simulations. The power ratios perform nearly as well on the simulated observations as on the exact models (§3.) for distinguishing multiclusters from single-component clusters. The point sources and Poisson noise do contribute noticeable uncertainty, but generally less than the systematic differences in power ratios between the two classes of models; the contribution to noise in the power ratios is roughly equal between the point sources and Poisson noise. Applying a Kolmogorov-Smirnov test to the power-ratios obtained from the simulations clearly demonstrates the sample of multiclusters may be distinguished from the sample of single-component clusters.

We applied the power-ratio method to ROSAT PSPC images of A85, A514, A1750, and A2029. These clusters each have very different X-ray morphologies with A2029 being a smooth, single-component cluster; A85 being a dominant smooth component with a small secondary; A1750 being two components of nearly equal size; and A514 being the quintessential complex cluster in the qualitative classification scheme of Forman & Jones (1990; Jones & Forman 1992); A85 and A1750 are also listed by Forman & Jones as the definitive members of their own classes. We find that the power ratios easily differentiate the clusters, especially when the aperture size is 1 Mpc ($H_0 = 80$ km s$^{-1}$ Mpc$^{-1}$). In a companion paper (Buote & Tsai 1995b) we apply the power ratios to a large sample of clusters observed with the PSPC.

We have discussed the suitability of the power-ratio method to constrain $\Omega$ via the Morphology - Cosmology connection (see §1.; RLT; Evrard et al. 1993). The method provides a simple, consistent comparison of the structure of clusters since the power ratios are (1) computed in a well-defined aperture, (2) normalized to the flux within that aperture, and (3) do not require any fitting. Moreover, the power-ratio method is specifically sensitive to the type of substructure described by RLT as relevant for cosmology. We also discussed the ability of the power ratio method to assess the viability of a particular cluster being described by hydrostatic equilibrium for the purposes of X-ray analysis of its intrinsic shape and of its total mass distribution.

It is a pleasure to thank Claude Canizares, Eric Gaidos, Lam Hui, and John Tonry for insightful discussions. We gratefully acknowledge Claude Canizares for a critical reading of the manuscript and Isamu Hatsukade for providing the A1750 data prior to public release. Finally, we thank Janet De Ponte at hotseat@cfa.harvard.edu for assistance in converting the German A1750 data to US/PROS format. DAB acknowledges grants NAS8-38249 and NASGW-2681 (through subcontract SVSV2-62002 from the Smithsonian Astrophysical Observatory). JCT was supported by an NRC associateship.

Table 1. Power Ratios

| Model | True | $z = 0.05$ | | $z = 0.10$ | | $z = 0.20$ | |
|---|---|---|---|---|---|---|---|

$P_2/P_0$ ($10^{-7}$)

Single ($\epsilon$):

| Model | True | $z = 0.05$ | | $z = 0.10$ | | $z = 0.20$ | |
|---|---|---|---|---|---|---|---|
| 0 | 0.000 | 0.160 | 0.068 - 0.182 | 0.482 | 0.181 - 0.548 | 1.61 | 0.609 - 1.86 |
| 0.3 | 39.6 | 41.9 | 38.4 - 45.4 | 42.5 | 35.6 - 49.4 | 43.2 | 29.9 - 56.8 |
| 0.6 | 162 | 167 | 160 - 175 | 169 | 155 - 183 | 168 | 141 - 196 |

Bimodal ($R_s, REL$):

| Model | True | $z = 0.05$ | | $z = 0.10$ | | $z = 0.20$ | |
|---|---|---|---|---|---|---|---|
| (0.5, 1 : 1) | 59.9 | 62.7 | 58.0 - 67.2 | 63.0 | 53.2 - 72.8 | 63.7 | 45.0 - 84.0 |
| (0.5, 2 : 1) | 46.3 | 48.0 | 43.7 - 52.1 | 49.1 | 40.5 - 57.7 | 48.7 | 32.7 - 66.1 |
| (0.5, 5 : 1) | 16.8 | 18.3 | 15.7 - 21.0 | 18.8 | 13.7 - 24.0 | 19.2 | 9.32 - 30.2 |
| (1.0, 1 : 1) | 882 | 865 | 845 - 887 | 876 | 834 - 918 | 860 | 707 - 945 |
| (1.0, 2 : 1) | 550 | 525 | 507 - 544 | 554 | 516 - 596 | 534 | 461 - 614 |
| (1.0, 5 : 1) | 85.9 | 82.2 | 75.1 - 89.1 | 90.9 | 77.1 - 107 | 90.5 | 58.9 - 122 |

$P_3/P_0$ ($10^{-7}$)

Single ($\epsilon$):

| Model | True | $z = 0.05$ | | $z = 0.10$ | | $z = 0.20$ | |
|---|---|---|---|---|---|---|---|
| 0 | 0.000 | 0.030 | 0.003 - 0.067 | 0.116 | 0.012 - 0.275 | 0.456 | 0.049 - 1.06 |
| 0.3 | 0.000 | 0.025 | 0.003 - 0.058 | 0.103 | 0.009 - 0.244 | 0.413 | 0.039 - 0.945 |
| 0.6 | 0.000 | 0.021 | 0.003 - 0.048 | 0.081 | 0.009 - 0.180 | 0.337 | 0.031 - 0.767 |

Bimodal ($R_s, REL$):

| Model | True | $z = 0.05$ | | $z = 0.10$ | | $z = 0.20$ | |
|---|---|---|---|---|---|---|---|
| (0.5, 1 : 1) | 0.000 | 0.032 | 0.004 - 0.075 | 0.133 | 0.013 - 0.320 | 0.501 | 0.054 - 1.14 |
| (0.5, 2 : 1) | 0.471 | 0.505 | 0.291 - 0.727 | 0.626 | 0.179 - 1.16 | 0.985 | 0.134 - 2.17 |
| (0.5, 5 : 1) | 0.664 | 0.693 | 0.441 - 0.960 | 0.805 | 0.299 - 1.38 | 1.18 | 0.196 - 2.47 |
| (1.0, 1 : 1) | 0.000 | 0.061 | 0.007 - 0.146 | 0.275 | 0.026 - 0.657 | 1.045 | 0.112 - 2.44 |
| (1.0, 2 : 1) | 29.6 | 28.4 | 26.2 - 30.6 | 28.7 | 24.2 - 33.2 | 28.6 | 20.4 - 37.3 |
| (1.0, 5 : 1) | 14.1 | 12.6 | 11.4 - 14.0 | 14.2 | 11.4 - 17.1 | 14.3 | 8.54 - 20.2 |

Table 1—Continued

| Model | True | $z = 0.05$ | | $z = 0.10$ | | $z = 0.20$ | |
|---|---|---|---|---|---|---|---|

$P_4/P_0$ ($10^{-7}$)

| Model | True | $z = 0.05$ | | $z = 0.10$ | | $z = 0.20$ | |
|---|---|---|---|---|---|---|---|
| Single ($\epsilon$): | | | | | | | |
| 0 | 0.000 | 0.014 | 0.001 - 0.033 | 0.054 | 0.006 - 0.121 | 0.209 | 0.022 - 0.474 |
| 0.3 | 0.099 | 0.135 | 0.068 - 0.205 | 0.170 | 0.042 - 0.324 | 0.320 | 0.046 - 0.703 |
| 0.6 | 2.18 | 2.24 | 1.97 - 2.51 | 2.34 | 1.76 - 2.95 | 2.53 | 1.43 - 3.76 |
| Bimodal ($R_s, REL$): | | | | | | | |
| (0.5, 1 : 1) | 0.036 | 0.068 | 0.021 - 0.120 | 0.107 | 0.014 - 0.233 | 0.303 | 0.037 - 0.700 |
| (0.5, 2 : 1) | 0.051 | 0.080 | 0.029 - 0.139 | 0.125 | 0.015 - 0.271 | 0.279 | 0.031 - 0.628 |
| (0.5, 5 : 1) | 0.056 | 0.087 | 0.031 - 0.150 | 0.131 | 0.020 - 0.281 | 0.273 | 0.032 - 0.610 |
| (1.0, 1 : 1) | 7.90 | 7.75 | 6.78 - 8.75 | 8.12 | 6.57 - 9.67 | 8.27 | 5.37 - 11.3 |
| (1.0, 2 : 1) | 10.1 | 9.65 | 8.84 - 10.4 | 10.1 | 8.36 - 11.7 | 9.94 | 6.67 - 13.3 |
| (1.0, 5 : 1) | 3.98 | 3.68 | 3.23 - 4.14 | 4.15 | 3.19 - 5.16 | 4.37 | 2.23 - 6.63 |

Note. — The power ratios are computed in a 1 Mpc circular aperture about the centroid. "True" corresponds to the intrinsic power ratio from §3. and the values for the different redshifts are the mean and 90% confidence limits for the 1000 simulated observations of the models (see §4.) having exposure time 10ks. The core radii and $\beta$ parameters of the models are fixed as described in §4.. The single-component models only differ in their ellipticity $\epsilon$. The bimodal multiclusters models are listed for different values of their relative separation, $R_s$ (Mpc), and their relative normalization, $REL$.

Table 2. Power Ratios

| Model | True | $z = 0.05$ | | $z = 0.10$ | | $z = 0.20$ | |
|---|---|---|---|---|---|---|---|

$P_1^{(pk)}/P_0^{(pk)}$ $(10^{-7})$

| Model | True | $z = 0.05$ | | $z = 0.10$ | | $z = 0.20$ | |
|---|---|---|---|---|---|---|---|
| Single ($\epsilon$): | | | | | | | |
| 0 | 0.000 | 2.36 | 0.078 - 3.31 | 5.21 | 0.370 - 8.43 | 14.8 | 1.64 - 29.6 |
| 0.3 | 0.000 | 0.798 | 0.065 - 1.49 | 2.87 | 0.239 - 6.68 | 11.6 | 1.10 - 25.9 |
| 0.6 | 0.000 | 0.588 | 0.051 - 1.27 | 2.47 | 0.207 - 3.77 | 8.17 | 0.758 - 19.3 |
| Bimodal ($R_s, REL$): | | | | | | | |
| (0.5, 1 : 1) | 4437 | 4378 | 3762 - 4851 | 3705 | 2887 - 4706 | 2896 | 1336 - 4192 |
| (0.5, 5 : 1) | 472 | 444 | 294 - 621 | 350 | 214 - 614 | 254 | 39.9 - 666 |
| (1.0, 1 : 1) | 5148 | 7092 | 6825 - 7318 | 5991 | 5574 - 6408 | 6149 | 5289 - 6955 |
| (1.0, 5 : 1) | 372 | 399 | 294 - 512 | 322 | 202 - 509 | 257 | 69.4 - 575 |

Note. — The power ratio is computed assuming an aperture of radius 1 Mpc centered on the emission peak. Quantities are listed as in Table 1

Table 3. Power Ratios vs. Aperture Size

| $R_{ap}$ (Mpc) | $P_1^{(pk)}/P_0^{(pk)}$ | $P_2/P_0$ | $P_3/P_0$ | $P_4/P_0$ |
|---|---|---|---|---|
| | | Single: $\epsilon = 0.3$: | | |
| 0.5 | 0. | 93.6 | 0. | 0.235 |
| 1.0 | 0. | 39.6 | 0. | 0.099 |
| 1.5 | 0. | 17.8 | 0. | 0.038 |
| 2.0 | 0. | 9.26 | 0. | 0.017 |
| | | Bimodal ($R_s = 1.0, REL = 2:1$): | | |
| 0.5 | 16.5 | 0.822 | 0.062 | 0.007 |
| 1.0 | 1824. | 550. | 29.6 | 10.1 |
| 1.5 | 3288. | 166. | 3.38 | 0.720 |
| 2.0 | 2078. | 55. | 0.643 | 0.083 |

Note. — The power ratios are expressed in units of $10^{-7}$.

Table 4: Observational Parameters

| Cluster | $z$ | Exposure (ks) | 0.1-2.4 keV Flux ($10^{-12}$ erg cm$^{-2}$ s$^{-1}$) | Background ($10^{-4}$ cts s$^{-1}$ arcmin$^{-2}$) |
|---|---|---|---|---|
| A85 | 0.0556 | 10.240 | 80.61 | 3.22 |
| A514 | 0.0731 | 18.111 | 5.00 | 2.24 |
| A1750 | 0.0855 | 13.148 | 14.62 | 2.79 |
| A2029 | 0.0768 | 12.550 | 66.67 | 5.50 |

Note. — Only the 0.5 and 1 Mpc values are listed for A85 because 1.5 Mpc lies outside the central ring of the PSPC. The fluxes are from Ebeling (1993) except A514 which we computed in this paper (see §5.). The background rate is computed in regions $\sim 30 - 40'$ from the field centers.

Table 5: Power Ratios of Abell Clusters

| | $P_1^{pk}/P_0^{pk}$ | | | $P_2/P_0$ | | | $P_3/P_0$ | | | $P_4/P_0$ | | |
|---|---|---|---|---|---|---|---|---|---|---|---|---|
| Cluster | 0.5 | 1.0 | 1.5 | 0.5 | 1.0 | 1.5 | 0.5 | 1.0 | 1.5 | 0.5 | 1.0 | 1.5 |
| A85 | 311 | 333 | ... | 15.4 | 13.8 | ... | 1.10 | 0.811 | ... | 0.032 | 0.166 | ... |
| A514 | 9369 | 4833 | 2557 | 273 | 300 | 179 | 22.7 | 5.83 | 7.40 | 0.679 | 13.6 | 0.155 |
| A1750 | 7 | 3670 | 3996 | 8.9 | 818 | 311 | 8.60 | 6.08 | 0.679 | 0.058 | 12.6 | 7.33 |
| A2029 | 37 | 9 | 5 | 14.0 | 1.7 | 2.0 | 0.031 | 0.004 | 0.020 | 0.050 | 0.073 | 0.060 |

Note. — Power ratios in units of $10^{-7}$ for real PSPC images of Abell clusters computed for aperture radii 0.5, 1, and 1.5 Mpc.

Table 6. Correlations of the Power Ratios

| Models | $P_1^{(pk)}/P_0^{(pk)}$ | | $P_2/P_0$ | | $P_3/P_0$ | | $P_4/P_0$ | |
|---|---|---|---|---|---|---|---|---|
| | 1.0 | 2.0 | 1.0 | 2.0 | 1.0 | 2.0 | 1.0 | 2.0 |
| Single ($\epsilon, a_x$): | | | | | | | | |
| (0.1, 0.1) | 0. | 0. | 0.527 | 0.081 | 0. | 0. | 7.8e-5 | 9e-6 |
| (0.1, 0.3) | 0. | 0. | 4.15 | 1.05 | 0. | 0. | 9.6e-4 | 1.9e-4 |
| (0.2, 0.1) | 0. | 0. | 2.07 | 0.314 | 0. | 0. | 1.3e-3 | 1.5e-4 |
| (0.2, 0.3) | 0. | 0. | 17.1 | 4.16 | 0. | 0. | 0.017 | 3.2e-3 |
| (0.3, 0.1) | 0. | 0. | 4.55 | 0.680 | 0. | 0. | 7.1e-3 | 8.0e-4 |
| (0.3, 0.3) | 0. | 0. | 39.6 | 9.26 | 0. | 0. | 0.099 | 0.017 |
| Bimodal ($\epsilon, R_s, REL$): | | | | | | | | |
| (0.0, 0.5, 1 : 1) | 4437 | 1223 | 59.9 | 4.44 | 0. | 0. | 0.036 | 2.3e-4 |
| (0.0, 0.5, 2 : 1) | 1929 | 543 | 46.3 | 3.50 | 0.471 | 0.010 | 0.051 | 3.3e-4 |
| (0.0, 0.5, 5 : 1) | 472 | 135 | 16.8 | 1.36 | 0.664 | 0.016 | 0.056 | 3.9e-4 |
| (0.0, 0.5, 10 : 1) | 139 | 40.3 | 5.68 | 0.480 | 0.330 | 8.3e-3 | 0.030 | 2.3e-4 |
| (0.0, 1.0, 1 : 1) | 5148 | 4716 | 882 | 70.6 | 0. | 0. | 7.90 | 0.059 |
| (0.0, 1.0, 2 : 1) | 1824 | 2078 | 550 | 55.3 | 29.6 | 0.643 | 10.1 | 0.089 |
| (0.0, 1.0, 5 : 1) | 372 | 515 | 85.9 | 21.1 | 14.1 | 0.968 | 3.98 | 0.096 |
| (0.0, 1.0, 10 : 1) | 101 | 153 | 17.5 | 7.31 | 3.47 | 0.502 | 0.993 | 0.055 |
| Major Axis: | | | | | | | | |
| (0.2, 0.5, 1 : 1) | 4367 | 1217 | 95.2 | 9.74 | 0.480 | 0.041 | 0.203 | 5.7e-3 |
| (0.2, 0.5, 2 : 1) | 1889 | 539 | 90.6 | 10.4 | 3.7e-3 | 6.4e-3 | 0.184 | 5.6e-3 |
| (0.2, 0.5, 5 : 1) | 460 | 134 | 56.7 | 8.19 | 0.179 | 1.5e-4 | 0.144 | 5.3e-3 |
| (0.2, 0.5, 10 : 1) | 135 | 39.9 | 37.6 | 6.48 | 0.117 | 5.6e-4 | 0.092 | 4.7e-3 |
| (0.2, 1.0, 1 : 1) | 5042 | 4690 | 993 | 88.2 | 1.15 | 0.152 | 12.5 | 0.155 |
| (0.2, 1.0, 2 : 1) | 1779 | 2062 | 677 | 76.9 | 19.1 | 0.202 | 13.0 | 0.159 |
| (0.2, 1.0, 5 : 1) | 362 | 510 | 163 | 39.3 | 10.3 | 0.577 | 4.55 | 0.144 |
| (0.2, 1.0, 10 : 1) | 98.6 | 151 | 64.6 | 20.7 | 2.57 | 0.330 | 1.25 | 0.086 |
| Minor Axis: | | | | | | | | |
| (0.2, 0.5, 1 : 1) | 4367 | 1217 | 36.4 | 1.25 | 0.629 | 0.046 | 1.9e-3 | 7.6e-5 |
| (0.2, 0.5, 2 : 1) | 1889 | 539 | 17.4 | 0.276 | 1.97 | 0.086 | 0.039 | 1.3e-3 |
| (0.2, 0.5, 5 : 1) | 460 | 134 | 0.401 | 0.285 | 1.53 | 0.059 | 0.099 | 3.7e-3 |
| (0.2, 0.5, 10 : 1) | 135 | 39.9 | 2.01 | 1.35 | 0.660 | 0.026 | 0.080 | 4.2e-3 |
| (0.2, 1.0, 1 : 1) | 5042 | 4690 | 850 | 56.6 | 3.19 | 0.186 | 5.49 | 0.020 |
| (0.2, 1.0, 2 : 1) | 1779 | 2062 | 423 | 37.7 | 47.1 | 1.43 | 9.12 | 0.062 |
| (0.2, 1.0, 5 : 1) | 362 | 510 | 28.6 | 8.35 | 17.6 | 1.49 | 4.09 | 0.112 |
| (0.2, 1.0, 10 : 1) | 98.6 | 151 | 0.026 | 0.708 | 4.29 | 0.716 | 1.19 | 0.078 |

Note. — The power ratios (in units of $10^{-7}$) computed in apertures of radii 1 Mpc and 2 Mpc; units of $a_x$ and $R_s$ are also Mpc and $\epsilon$ in the bimodal models refers to the ellipticity of the primary component. Bimodal models where the secondary component lies along the major axis of the primary are listed under the heading "Major Axis"; similarly the models where the secondary is aligned along the minor axis are listed under "Minor Axis".

Fig. 1.—
Contour plots of four simple models for X-ray clusters placed at $z = 0.10$: (a) is a single-component model with core radius 300 kpc and $\epsilon = 0.30$; (b) - (d) are bimodals each separated by 1 Mpc and have core radii 300 kpc but the clumps are in proportion 1 : 1 for (b), 2 : 1 for (c), and 5 : 1 for (d). The contours are separated by factors of two in surface brightness (arbitrary units). The units are in $15''$ pixels which translates to 36.7 pixels/Mpc and 2.7 Mpc/side for $H_0 = 80$ km/s/Mpc.

Fig. 2.—
Contour plots of 10ks simulated observations of the cluster models in Figure 1 as described in §4.. The images have been smoothed with the PSPC PSF for viewing purposes only.

Fig. 3.—
Cumulative distributions of $P_1^{(pk)}/P_0^{(pk)}$, $P_2/P_0$, $P_3/P_0$, and $P_4/P_0$ for the simulated cluster samples of single-component clusters (solid) and bimodal clusters (dotted).

Fig. 4.—
Contour plots of the PSPC images of Abell clusters A85, A514, A1750, and A2029 corrected for exposure, vignetting, and background; the angular sizes of the fields are the same as Figure 1. The contours are separated by factors of 2 in intensity and the images have been smoothed with the PSPC PSF for viewing purposes only.

Fig. 5.—
Contour plots of the PSPC images of Abell clusters A85, A514, A1750, and A2029 prepared as in Figure 4 but with sources removed as described in §5.. The images have been smoothed with the PSPC PSF for viewing purposes only.

Fig. 6.—
Correlations of the power ratios for a sample of models of single-component clusters (filled ovals) and bimodal multiclusters (crosses); see Table 3 for a description of the sample. The dashed lines represent the region where single-component clusters are allowed due to observational uncertainty as determined by the simulated PSPC observations in §4..

Fig. 7.—
Same as Figure 6 for the power ratios computed in the 2 Mpc aperture.